\documentclass[12pt]{article}

\usepackage{amssymb}

\begin{document}




\title{A Common Thread\footnote{Published in Physica C: Superconductivity doi:10.1016/j.physc.2010.01.004}}


\author{Douglas J. Scalapino \\ University of California, Physics Department,\\ Santa Barbara, CA 93106-9530, USA}






\maketitle

\begin{abstract}
The structures, the phase diagrams, and the appearance of a neutron resonance
in the superconducting state provide phenomenological evidence which relate the
heavy fermion, cuprate and Fe superconductors. Single- and multi-band Hubbard
models have been found to describe a number of the observed properties of
these materials so that it is reasonable to examine the origin of the pairing
interaction in these models. Here based on the experimental phenomenology and
studies of the momentum and frequency dependence of the pairing interaction for
Hubbard-like models, we suggest that spin-fluctuation mediated pairing is the
common thread linking this broad class of superconducting materials. 

\end{abstract}




For a number of years there have been suggestions that just as the electro-phonon
interaction leads to pairing in a variety of non-magnetic superconducting
materials, pairing mediated by the exchange of spin-fluctuations is responsible
for the superconducting seen in a wide class of nearly antiferromagnetic
materials. These suggestions have ranged from proposals that spin-fluctuations
near an antiferromagnetic instability were responsible for pairing in some
organic BEDT salts and heavy fermion systems \cite{ref:1,ref:2} to later work
that included the cuprate \cite{ref:3,ref:4,ref:5} as well as the recently
discovered Fe superconductors \cite{ref:6,ref:7,ref:G}. One could also include
Sr$_2$RuO$_4$ in this class of unconventional superconductors \cite{ref:8}, where in this case ferromagnetic spin-fluctuations are important.
Here we will focus on the heavy fermion, cuprate and Fe superconductors. We
first argue that their chemical and structural makeup, their phase diagrams and
evidence of quantum critical behavior, along with the observation of a neutron
scattering spin resonance in the superconducting phase support the notion that
they form a related class of superconducting materials. We then note that a
number of their observed properties are described by Hubbard-like models and
that calculations show that the pairing in such models arises from the exchange
of spin fluctuations. We conclude that pairing arising from the exchange of
spin-fluctuations near an antiferromagnetic instability is the
common thread linking the heavy fermion, cuprate and Fe superconductors\cite{ref:MO}.

To begin with, these materials are all known to occur in families.
For example, one has the 115 heavy fermion family CeTIn$_5$ with T=Co, Rh
and Ir, and the actinide 115 PuTGa$_5$ with T=Co and Rh. For the cuprates,
one has the well-known Hg, Tl and Bi families and similarly for the Fe
superconductors there are a number of different families including the
(1111) LaOFeAs, the 122 Ba(FeAs)$_2$ and the (11) Fe(Se$_{1-x}$Te$_x$) groups.

Structurally, these materials are layered quasi-2D systems with a square planar
arrangement of ions containing hybridized d or f orbitals which have local
magnetic character and lay near the
Fermi energy. In the heavy fermion materials, the f-electrons hybridize with the
spd electrons and the parent compound is a low temperature, coherent, heavy mass,
paramagnetic metal. Pressure or doping can alter the delicate balance between
antiferromagnetism and superconductivity. In the cuprates, the Cu 3d electrons
hybridize with O p-orbitals and the parent compound is a charge-transfer
antiferromagnetic Mott insulator which becomes superconducting when doped.
In the Fe-pnictide and chalcogen families, the Fe 3d orbitals hybridize through
As or Se (Te) 4p orbitals and the parent compound is a semi-metallic
antiferromagnet. In this case, pressure, chemical pressure or doping can lead
to superconductivity.

The phase diagrams of these materials typically exhibit antiferromagnetism
near or in some cases coexisting with superconductivity. Furthermore, via
doping, chemical pressure or hydrostatic pressure one can move from the
antiferromagnetic to the superconducting phase. In addition, there can also
occur nematic as well as
striped magnetic and charge density wave phases. In some cases, such as the
underdoped cuprates, one is near a charge-transfer antiferromagnetic Mott
phase. However, the common element seen in the various phase diagrams of the
heavy-fermion \cite{ref:9,ref:10,ref:11}, cuprate \cite{ref:12,ref:13} and Fe
\cite{ref:14,ref:15} families is the proximity of the antiferromagnetic
and superconducting regions.

Another common feature is evidence for the occurrence of a quantum critical
point (or points), veiled by the superconducting dome, which gives rise to
characteristic power law temperature dependencies at temperatures above $T_c$
\cite{ref:15a,ref:15b,ref:15c,ref:15d}. Along with this, there can be
coexistence of magnetic and superconducting order and electronic structure
reconstruction due to the presence of the magnetic order \cite{ref:C,ref:D,ref:H}.
In all three classes
of materials there is evidence of a sensitivity of the superconducting pairing
to the structure of the Fermi surface or surfaces.

In these materials, the ratio of the superconducting transition temperature $T_c$
to the Fermi energy or coherence scale is similar and large relative to that of
the traditional electron-phonon superconductors \cite{ref:4}. Furthermore,
electronic structure frozen phonon calculations of the electron-phonon pairing
strength find that it is far too weak to give the observed superconducting
transition temperatures of these systems \cite{ref:16,ref:17}.

Finally, a key experimental observation linking these materials is the appearance
of a neutron scattering magnetic resonance in the superconducting state at an
antiferromagnetic or SDW wave vector $Q$. This type of resonance, first observed
in YBCO by Rossat-Mignod \cite{ref:18}, has now been seen in the heavy-fermion
\cite{ref:19}, cuprate \cite{ref:20,ref:21} and Fe \cite{ref:22,ref:23}
superconductors. It implies that the gap changes sign between regions on the
Fermi surface (or surfaces) separated by momentum $Q$, which contribute
significantly to the spin scattering \cite{ref:24,ref:25}
\begin{equation}
  \Delta(k+Q)=-\Delta(k).
\end{equation}
Thus, all of these materials have an ``unconventional" gap with a magnetic
resonance at an energy which scales with the gap \cite{ref:25a}. In addition,
the magnitude of the gap typically rises rapidly as the temperature
decreases below $T_c$ and reaches a $2\Delta_0/kT_c$ value which is larger than
the expected BCS value. This behavior is consistent with a picture in which not
only are the same electrons involved with both the magnetic and superconducting
properties, but the same electrons that are pairing participate in the pairing
interaction \cite{ref:24,ref:25b}.

From a theoretical perspective, the single- and multi-band Hubbard models
provide a framework for describing these materials \cite{ref:26,ref:27}. While
a multi-band Hubbard model is needed to describe the charge-transfer aspect of
the cuprates and 5-band models are presently used for the Fe-superconductors,
a substantial part of the basic underlying physics is believed to be captured in
the single-band Hubbard model. For example, this basic model at half-filling is
known to exhibit an antiferromagnetic Mott insulating state, while the doped
system exhibits stripes and d-wave pairfield correlations \cite{ref:28}.
There are also numerical calculations which show that there is a magnetic
resonance peak present in the superconducting state \cite{ref:29}.

Calculations of the irreducible particle-particle pairing interaction for the
single band Hubbard model show that
it is associated with the particle-hole $S=1$ exchange channel, i.e.,
spin-fluctuations \cite{ref:30}. In particular, recent calculations show that
the dynamics of the pairing interaction reflect the structure in the imaginary
part of the dynamic spin susceptibility \cite{ref:31}. There have also been a
number of RPA \cite{ref:7,ref:G} and weak coupling renormalization
group treatments \cite{ref:32,ref:33} of 5-band Hubbard models for Fe which
conclude that spin-fluctuations provide the pairing interaction in the
Fe-pnictides and chalcogen superconductors.

Beyond the pairing interaction, which we take to be the irreducible 4-point
vertex in the zero center of mass momentum and energy particle-particle
channel, there are a number of other factors which influence the onset of
superconductivity. There is of course the structure of the single particle
propagator. Near a Mott-antiferromagnetic transition, the superconducting
transition temperature is suppressed by the Mott gap which opens in the single
particle spectrum even though the 4-point vertex pairing interaction is strong.
There is the interplay of competing orders, the role of nematic and stripe-like
fluctuations \cite{ref:AA} and the possible reconstruction or fluctuation of the
Fermi surface \cite{ref:BB, ref:34}. There is also the interesting possibility that an
optimally inhomogenous structure may lead to an enhanced $T_c$ \cite{ref:CC}.
Nevertheless, from calculations carried out on clusters with inhomogeneities it
appears that the short range spin-fluctuations mediate the pairing interaction \cite{ref:35}.

In summary, there are similar material, experimental and theoretical features
relating the heavy fermion, cuprate and Fe superconductors. The materials come
in families which contain quasi-2D layers of correlated d or f electrons.
Their temperature-doping and magnetic field phase diagrams show
antiferromagnetism in close proximity to superconductivity.
The resonant peak observed in inelastic neutron scattering experiments in the
superconducting phase provide a signature for an unconventional gap
$\Delta(k+Q)=-\Delta(k)$. This resonance also implies that the same electrons
are involved in both the magnetism and the superconductivity. The rapid onset
of the gap as the temperature is lowered below $T_c$ as well as the large
$2\Delta_0/kT_c$ ratio further suggest that these electrons are also
involved in the effective pairing interaction.

Single- and multi-band Hubbard models exhibit a number of the properties seen
in these materials. Numerical studies of the effective pairing interaction in
the single-band Hubbard model and various weak coupling calculations on
multi-band models find unconventional pairing mediated by an $S=1$
particle-hole channel. 
Thus while the heavy fermion, cuprate and Fe-pnictide (or chalcogen) materials
exhibit a wide range of properties, we believe that spin-fluctuated mediated
pairing provides the common thread which is responsible for superconductivity
in all of these materials.

\section*{Acknowledgment}

This work arose from an invitation to present an overview regarding "the
mechanism of high temperature superconductivity including the cuprates and Fe-
pnictides" at the 9th International Conference on Materials and Mechanisms of
Superconductivity held in Tokyo (September 7--12, 2009). Here I have broadened
the materials to include the heavy-fermions and narrowed the mechanism question
to focus simply on the pairing interaction. I want to thank many coworkers,
some of whom are listed in the references, as well as many experimentalists
including Z.~Fisk, B.~Keimer, J.~Thompson and L.~Taillefer for sharing their insights.
I would also like to acknowledge the support of the Center for Nanophase Materials Science at
ORNL, which is sponsored by the Division of Scientific User Facilities, U.S.\ DOE.






\begin{thebibliography}{00}


\bibitem{ref:1} K.~Miyaki, S.~Schmitt-Rink and C.M.~Varma, {\it Phys.\ Rev.\ B}
{\bf 34}, 6554 (1986).
\bibitem{ref:2} D.J.~Scalapino, E.~Loh, Jr.\ and J..E.~Hirsch, {\it Phys.\ Rev.\ B}
{\bf 34}, 8190 (1986).
\bibitem{ref:3} D.J.~Scalapino, {\it Phys.\ Rep.} {\bf 250}, 329 (1995).
\bibitem{ref:4} T.~Moriya and K.~Ueda, {\it Rep.\ Prog.\ Phys.} {\bf 66}, 1299 (2003).
\bibitem{ref:5} P.~Monthoux, D.~Pines and G.~Longarich, {\it Nature} {bf 450}, 20 (2007).
\bibitem{ref:6} I.I.~Mazin, D.J.~Singh, M.D.~Johannes and M.H.~Du,
{\it Phys.\ Rev.\ Lett.} {\bf 101}, 057003 (2008).
\bibitem{ref:7} K.~Kuroki, S.~Onari, R.~Arita, H.~Usui, Y.~Tanaka,
H.~Kontani and H.~Aoki, {\it Phys. Rev. Lett.} {\bf 101}, 087004 (2008).
\bibitem{ref:G} S.~Graser, T.A.~Maier, P.J.~Hirschfeld and D.J.~Scalapino,
New J.\ Phys. {\bf 11} 025016 (2009).
\bibitem{ref:8} A.P.~Mackenzie and Y.~Maeno, {\it Rev. Mod. Physics} {\bf 75}, 657 (2003).
\bibitem{ref:MO} This of course is one view of what remains one of the important questions in condensed matter physics. As is well known, there are a number of different proposals for the mechanism responsible for high Tc superconductivity. In particular, carrier doping into a Mott insulating state is seen as the key feature in some approaches. For a recent review which takes this point of view, see M. Ogata and H. Fukuyama, Rep. Prog. Phys. 71,036501 (2008).
\bibitem{ref:9} N.D.~Mathur, F.M.~Grosche, S.R.~Julian, I.R.~Walker, D.M.~Freye,
R.K.W.~Haselwimmer and G.G.~Lonzarich, {\it Nature} {\bf 394}, 39 (1998).
\bibitem{ref:10} N.~Nicklas, O.~Stockert, T.~Park, K.~Habicht, K.~Kiefer,
L.D.~Pham, J.D.~Thompson, Z.~Fisk, and F.~Steglich, {\it Phys.\ Rev.\ B} {\bf 76}, 052401 (2007).
\bibitem{ref:11} L.D.~Pham, T.~Park, S.~Maquilon, J.D.~Thompson and Z.~Fisk, {\it Phys.\ Rev.\ Lett.} {\bf 97}, 056404 (2006).
\bibitem{ref:12} N.E.~Hussey, {\it J. Phys: Condens. Matter} {\bf 20}, 123201 (2008).
\bibitem{ref:13} C.~Panagopoulos, A.P.~Petrovic, A.D.~Hillier, J.L.~Tallon,
C.A.~Scott and B.D.~Rainford, {\it Phys.\ Rev.\ B} {\bf 69}, 144510 (2004).
\bibitem{ref:14} J.~Zhao, Q.~Huang, C.~de la Cruz, S.~Li, J.W.~Lynn, Y.~Chen,
M.A.~Green, G.F.~Chen, G.~Li, Z.~Li, J.L.~Luo, N.L.~Wang and P.~Dai, {\it Nature Materials} {\bf 7}, 953 (2008).
\bibitem{ref:15} H.~Chen, Y.~Ren, Y.~Qiu, W.~Bao, R.H.~Liu, G.~Wu, T.~Wu,
Y.L.~Xie, X.F.~Wang, Q.~Huang and X.H.~Chen, {\it Europhys.\ Lett.} {\bf 85}, 17006 (2009).
\bibitem{ref:15a} T.~Park, Y.~Tokiwa, F.~Ronning, H.~Lee, E.D.~Bauer,
R.~Movshovich and J.D.~Thompson, arXiv:0910.2287
\bibitem{ref:15b} S.E.~Sebastian, N.~Harrison, M.M.~Altarawneh, C.H.~Mielke,
R.~Liang, D.A.~Bonn, W.N.~Hardy and G.G.~Lonzarich, arXiv:0910.2359
\bibitem{ref:15c} C.~Liu, T.~Kondo, R.M.~Fernandes, A.D.~Palczewski, E.D.~Mun, N.~Ni,
A.N.~Thaler, A.~Bostwick, E.~Rotenberg, J.~Schmalian, S.L.~Bud'ko, P.C.~Canfield and A.~Kaminski, arXiv:0910.1799
\bibitem{ref:15d} The organic superconductor (TMTSF)$_2$PF$_6$ has a similar
phase diagram and linear $T$ resistivity as the overdoped cuprates
N.~Doiron-Leyraud, P.~Auban-Senzier, S.~Rene~de~Cotret, A.~Sedeki, C.~Bourbonnais,
D.~Jerome, K.~Bechgaard, and L.~Taillefer, arXiv:0905.0964
\bibitem{ref:C} J.~Chang, R.~Daou, C.~Proust, D.~LeBoeuf, N.~Doiron-Leyraud,
F.~Laliberte, B.~Pingault, B.J.~Ramshaw, R.~Liang, D.A.~Bonn, W.N.~Hardy, H.~Takagi,
A.~Antunes, I.~Sheikin, K.~Behnia and L.~Taillefer, arXiv:0907.5039
\bibitem{ref:D} N.~Doiron-Leyraud, C.~Proust, D.~LeBoeuf, J.~Levallois,
J.-B.~Bonnemaison, R.~Liang, D.A.~Bonn, W.N.~Hardy and L.~Taillefer, {\it Nature} {\bf 447}, 565 (2007).
\bibitem{ref:H} N.~Harrison, R.D.~McDonald, C.H.~Mielke, E.D.~Bauer, F.~Ronning
and J.D.~Thompson, arXiv:0902.1481.
\bibitem{ref:16} S.Y.~Savrasov and O.K.~Andersen, {\it Phys.\ Rev.\ Lett.} {\bf 71},
4430 (1996).
\bibitem{ref:17} L.~Boeri, O.V.~Dolgov and A.A.~Golubov, {\it Phys.\ Rev.\ Lett.}
{\bf 101}, 026403 (2008).
\bibitem{ref:18} J.~Rossat-Mignod, L.P.~Regnault, C.~Vettier, P.~Bourges,
P.~Burlet, J.~Bossy, J.Y.~Henry and G.~Lapertot, {\it Physica C} {\bf 185}, 86 (1991).
\bibitem{ref:19} C.~Stock, C.~Broholm, J.~Hudis, H.J.~Kang and C.~Petrovic,
{\it Phys.\ Rev.\ Lett.} {\bf 100}, 87001 (2008).
\bibitem{ref:20} H.A.~Mook, M.~Yethiraj, G.~Aeppli, T.E.~Mason and T.~Armstrong,
{\it Phys.\ Rev.\ Lett.} {\bf 70}, 3490 (1993).
\bibitem{ref:21} H.F.~Fong, B.~Keimer, P.W.~Anderson, D.~Reznik, F.~Dogan and
I.A.~Aksay, {\it Phys.\ Rev.\ Lett.} {\bf 75}, 316 (1995).
\bibitem{ref:22} A.D.~Christiansen, E.A.~Goremychkin, R.~Osborn, S.~Rosenkranz,
M.D.~Lumsden, C.D.~Malliakas, I.S.~Todorov, H.~Claus, D.Y.~Chung, M.G.~Kanatzidis,
R.I.~Bewley and T.~Guidi, {\it Nature} {\bf 456}, 930 (2008).
\bibitem{ref:23} D.S.~Inosov, J.T.~Park, P.~Bourges, D.L.~Sun, Y.~Sidis,
A.~Schneidewind, K.~Hradil, D.~Haug, C.T.~Lin, B.~Keimer and V.~Hinkov, arXiv:0907.3632
\bibitem{ref:24} P.~Monthoux and D.J.~Scalapino, {\it Phys.\ Rev.\ Lett.} {\bf 72},
1874 (1994).
\bibitem{ref:25} C.-H.~Pao and N.E.~Bickers, {\it Phys.\ Rev.\ B} {\bf 51}, 16310 (1995).
\bibitem{ref:25a} G.~Yu, Y.~Li, E.M.~Motoyama and M.~Greven, {\it Nature Physics}
arXiv:0903.2291
\bibitem{ref:25b} C.-H.~Pao and N.E.~Bickers, {\it Phys.\ Rev.\ Lett.} {\bf 72},
1870 (1994).
\bibitem{ref:26} P.W.~Anderson, {\it Science} {\bf 235}, 1196 (1987).
\bibitem{ref:27} T.~Takimoto, T.~Hotta and K.~Ueda, {\it Phys. Rev. B} {\bf 69},
104504 (2004).
\bibitem{ref:28} For a review see D.J.~Scalapino, {\it Handbook of High Temperature
Superconductivity}, Chapter 13, Eds.\ J.R.~Schrieffer and J.S.~Brooks (Springer,
New York, 2007). 
\bibitem{ref:29} S.~Brehm, E.~Arrigoni, M.~Aichhorn and W.~Hanke, arXiv:0811.0552
\bibitem{ref:30} T.A.~Maier, M.~Jarrell and D.J.~Scalapino,	{\it Phys. Rev.\ B}
{\bf 74}, 94513 (2006).
\bibitem{ref:31} B.~Kyung, D.~Senechal and A.-M.S.~Tremblay, arXiv:0812.1228
\bibitem{ref:32} F.~Wang, H.~Zhai, Y.~Ran, A.~Vishwanath, and D.H.~Lee,
{\it Phys.\ Rev.\ Lett.} {\bf 102}, 047005 (2009).
\bibitem{ref:33} C.~Platt, C.~Honerkamp and W.~Hanke, {\it New J.\ Phys.} {\bf 11},
055052 (2009).
\bibitem{ref:AA} S.A.~Kivelson, I.P.~Bindloss, E.~Fradkin, V.~Oganesyan,
J.~Tranquada, A.~Kapitulnik and C.~Howald, {\it Rev.\ Mod.\ Phys.} {\bf 75}, 1201 (2008).
\bibitem{ref:BB} S.~Sachdev, M.A.~Metlitski, Y.~Qi and C.~Yu, {\it Phys.\ Rev.\ B} {\bf 80},
155129 (2009).
\bibitem{ref:34} S.~Sachdev, arXiv:0907.0008, {\it Physica Status Solidi},
29 September (2009).

\bibitem{ref:CC} E.~Berg, D.~Orgad and S.A.~Kivelson, {\it Phys.\ Rev.\ B} {\bf 78}, 094509 (2008).

\bibitem{ref:35} T.A.~Maier, G.~Alvarez and T.C.~Shulthess, in preparation.

\end{thebibliography}



\end{document}